\documentclass[12pt]{article}
\usepackage[dvips]{graphicx}
\usepackage{amsfonts}
\usepackage{graphics}

\newtheorem{prop}{Proposition}[section]

\title{Quantum Computation \\ Beyond the "Standard Circuit Model"
\footnote{submitted to the proceedings of the NATO-ASI conference
on Quantum Information and Quantum Computation, May 2005, Chania,
Greece.}}

\author{~K.~Ch.~Chatzisavvas$^{a}$\footnote{\texttt{e-mail:\, kchatz\,@\,auth.gr}}\,,
        ~C.~Daskaloyannis$^{b}$\footnote{\texttt{e-mail:\, daskalo\,@\,auth.gr}}\,,
        ~C.~P.~Panos$^{a}$\footnote{\texttt{e-mail:\,
        chpanos\,@\,auth.gr}}
\\
 {\it  $^{a}$Department of Theoretical Physics,}\\
 {\it $^{b}$Department of Mathematics,}\\
        {\it Aristotle University of Thessaloniki,}\\
                {\it  54124 Thessaloniki, Greece} \\
 }

\date{}
\begin{document}

\maketitle

\begin{abstract}

Construction of explicit quantum circuits follows the notion of
the "standard circuit model" introduced in the solid and profound
analysis of elementary gates providing quantum computation.
Nevertheless the model is not always optimal (e.g. concerning the
number of computational steps) and it neglects physical systems
which cannot follow the "standard circuit model" analysis. We
propose a computational scheme which overcomes the notion of the
transposition from classical circuits providing a computation
scheme with the least possible number of Hamiltonians in order to
minimize the physical resources needed to perform quantum
computation and to succeed a minimization of the computational
procedure (minimizing the number of computational steps needed to
perform an arbitrary unitary transformation). It is a general
scheme of construction, independent of the specific system used
for the implementation of the quantum computer. The open problem
of controllability in Lie groups is directly related and rises to
prominence in an effort to perform universal quantum computation.
\end{abstract}


\section{The "Standard Circuit Model"}

The "standard circuit model" is an 
established proposal to implement quantum gates in quantum
computation \cite{Barenco95}. In this model essential is the
notion of the universal gate \cite{Deutsch85}. Thus, any given
quantum gate (any given unitary transformation of the quantum
system that implements the quantum computer) can be analyzed using
a set of basic gates, known as universal gates. The selection of
the set of universal gates is not unique \cite{Lloyd95}. One-qubit
gates can be analyzed using only Hadamard and phase gates.
Two-qubit gates can be analyzed using Hadamard, phase and the CNOT
gate and this is generalized in the case of $N$-qubit gates, while
it was noted that in the general case an infinite number of steps
are needed to perform a gate explicitly \cite{Deutch95}.

In the "standard circuit model", physical systems are neglected if
they cannot copy easily the model (if someone cannot perform
easily one of the selected universal gates). Also, neither the
number of computational steps nor the total time to perform
computation are optimal \cite{Sproel05}.

\section{Quantum Control Theory}
In quantum control theory, the generalization of the control
theory in quantum systems, a system is said to be controllable if
an arbitrary Lie group element $W\in SU(2^N)$ can be decomposed in
finite time as
\begin{equation}
W=e^{-i a_n J^{(n)}}\,\ldots\,e^{-i a_2 J^{(2)}}\,e^{-i a_1
J^{(1)}}
\end{equation}
where $J^{(k)}\in\{J_1,J_2,\ldots,J_m\}$ are generators of the
corresponding $su(2^N)$ Lie algebra and $a_i \in\mathbb{R}$. In
the case of quantum computation, $W$ is equivalent with an
arbitrary unitary transformation (up to a global phase) so it is
equivalent with an arbitrary $N$-qubit gate. $J^{(k)}$ corresponds
to the Hamiltonians describing the system under consideration
while $a_i$ are equivalent with time parameters $t_i$. The
controllability on Lie groups from a mathematical point of view
was studied in \cite{JurSus75,Rama95,Khaneja01,SchFuSo01}. This
direct relation between the problem of controllability in Lie
groups and the problem of universal quantum computation allows us
to approach quantum computation with an alternative way beyond the
"standard circuit model". In this approach if the selected
Hamiltonians $J_1,\, J_2,\, \ldots, J_m$ form a complete set of
operators, then every $W\in SU(2^N)$ can be exactly realized using
a finite number of steps, although this number of steps is not
fixed, where in the case of the "standard circuit model" the same
element  $SU(2^N)$ could be approximately realized using an
infinitely  number of steps. The order of generation (the number
of computational steps required to perform an arbitrary N-qubit
gate) is available for arbitrary Hamiltonians only in the case of
the $SU(2)$ group (one-qubit gates) via the Lowental's criterion
\cite{Lowenthal72}. In this case, only two Hamiltonians
$\left\{J_1,\, J_2\right\}$ are sufficient to form a complete set.
If the Hamiltonians are orthogonal, i.e. ${\rm Trace}(J_1 J_2)=0$,
then three at most steps are required, to realizing any $W\in
SU(2)$). When ${\rm Trace}(J_1 J_2)\ne 0$, the number of steps are
given by the Lowental's criterion, but the algorithm to obtain the
solution is not known.

In the case of higher order groups there is an analysis based on
the Cartan decomposition of the $su(2^N)$ algebras
\cite{Khaneja01}. This analysis provides also an analytical way of
calculating the corresponding time parameters (Euler angles) in
the case of the $SU(4)$ (2-qubit gates). On the same spirit is the
proposal for exact computation by Whaley and collaborators
\cite{Zhang03}. Open problems in Lie groups controllability are:
\begin{itemize}
\item[a)] \emph{SU(2) group (one-qubit gates)}. An algorithm
which, given an arbitrary couple of generators--Hamiltonians, will
be able to provide analytically the time parameters to perform
universal computation, if the number of required steps are more
than three.
 \item[b)] \emph{Higher order groups}. A
criterion for minimum number of steps to generate an arbitrary
element of the group (which corresponds to an arbitrary N-qubit
gate, respectively) in the case where the generators--Hamiltonians
are not orthogonal. Algorithms to evaluate the corresponding time
parameters.
\end{itemize}

\section{Quantum Gates Using the Intrinsic Abilities of a Physical System}

The main points of our proposal can be summarized briefly as
follows
\begin{itemize}
\item Instead of forcing a physical system to act as a
predetermined set of universal gates we focus on the ability of
the physical system to act as a quantum computer using only its
natural available interactions (encoded universality
\cite{DiVincenzo00}). \item Construction of any given gate and
algorithm in terms of a minimal configuration and computational
procedure. \item Minimized finite number of steps, evolving in
time according to a finite number of basic, intrinsic
Hamiltonians, controlled by a minimal finite number of classical
switches (the selection of the switches is not unique). \item
Implementation does not depend on the psecific system used as a
qubit. Several solid state proposals as charge Josephson
junctions, SQUIDs, quantum dots have been tested but our proposal
can be extended to NMR quantum computation, trapped ions etc in
order to test it with various physical systems described by
different Hamiltonians and interactions.
\end{itemize}

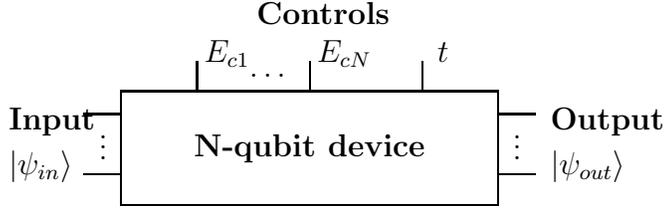
\begin{figure}[h]
\begin{center}
\begin{picture}(10,3.5)
\put(2.5,1.0){\framebox(5.0,1.5){\textbf{N-qubit device}}}
\put(1.0,2.0) {\textbf{Input}}
\put(1.0,1.4){{$|\psi_{in}\rangle$}}

\put(2.5,1.4){\line(-1,0){0.5}} \put(2.5,2.2){\line(-1,0){0.5}}
\put(2.2,1.6){\vdots}

\put(8.2,2.0) {\textbf{Output}}
\put(8.2,1.4){{$|\psi_{out}\rangle$}}

\put(7.5,1.4){\line(1,0){0.5}} \put(7.5,2.2){\line(1,0){0.5}}
\put(7.7,1.6){\vdots}

\put(3.5,2.5){\line(0,1){0.4}}\put(3.6,2.9){$E_{c1}$}
\put(5.0,2.5){\line(0,1){0.4}}\put(5.1,2.9){$E_{cN}$}
\put(6.5,2.5){\line(0,1){0.4}}\put(6.7,2.9){$t$}
\put(4.2,2.7){\ldots} \put(4.3,3.4){\textbf{Controls}}
\end{picture}
\end{center}
\caption{\it{Abstract N-qubit device}}\label{fig:1}
\end{figure}

This computer consists of one cell controlled by external binary
switches and evolving in time using these switches. Quantum gates
and algorithms are translated into manipulation of these switches.
It is a simple device which overcomes the notion of transposition
from classical circuits and does not have any "quantum"
connections (one of the difficult parts in physical
implementation-especially in solid state devices).

The above proposal is based in the following mathematical
Propositions:

\begin{prop}\label{prop:switch}
A number of $N+1$ switches are sufficient for universal quantum
computation in a $N$-qubit device.
\end{prop}

\begin{prop}\label{prop:Hamilt}
A set of $N+2$ Hamiltonians provided by the basic Hamiltonian of
the $N$-qubit device through appropriate tuning of the $N+1$
switches, can generate the $su(2^N)$ algebra.
\end{prop}

\begin{prop}\label{prop:construction}
The construction scheme of any quantum gate consists of a finite
number of steps evolving in time according to a finite number of
basic Hamiltonians (reiterating in cyclic pattern) and  provided
by proper switches' manipulation.
\begin{equation}
  U=\ldots \textrm{e}^{-i t_{4} H_{4}}\textrm{e}^{-i t_{3} H_{2}}
    \textrm{e}^{-i t_{2} H_{3}} \textrm{e}^{-i t_{1} H_{1}}
\end{equation}
It is an open conjecture that every $U$ can be generated exactly
by $\mathcal{O}(4^N)$ steps.
\end{prop}

The manipulations  of the quantum computer can be codified by a
rudimentary \emph{Quantum Machine Language} \cite{Chatzisavvas03}.

\section{Results}

\subsection{One-qubit gates}
According to the "standard model" the most usual analysis of an
arbitrary one-qubit gate includes two Hadamard gates and two phase
gates.
\begin{center}
\begin{picture}(13.0,1.0) \thicklines
\put(1.5,0.0){\framebox(1,1){$W$}}\put(1.5,0.5){\line(-1,0){0.5}}
\put(2.5,0.5){\line(1,0){0.5}} \put(3.6,0.4){$=$}
\put(4.5,0.5){\line(1,0){0.5}}\put(5.0,0.0){\framebox(1,1){$H$}}\put(6.0,0.5){\line(1,0){0.25}}
\put(6.25,0.0){\framebox(2.0,1.0){$2\theta$}}
\put(8.25,0.5){\line(1,0){0.25}}\put(8.5,0.0){\framebox(1.0,1.0){$H$}}
\put(9.5,0.5){\line(1,0){0.25}}\put(9.75,0.0){\framebox(2.0,1.0){$\frac{\pi}{2}+\phi$}}
\put(11.75,0.5){\line(1,0){0.5}}
\end{picture}
\end{center}
Even if both Hadamard and phase gate can be realized in one
computational step, then at least four steps are required to
perform universal quantum computation. Usually decompositions of
elementary one-qubit gates require at least 8 computational steps
according to this analysis and it is performed most of the times
in systems which provide us with orthogonal
Hamiltonians. \\

With the presented computational scheme the results in the case of
the one-qubit gates are the following:

\noindent \textbf{Orthogonal Hamiltonians} If the Hamiltonians are
orthogonal, i.e. $(H_1,H_2)={\rm Trace}(H_1 H_2)=0$, then two
Hamiltonians and three computational steps at most are required,
to realize any $W\in SU(2)$.
\begin{equation}
W=e^{-i t_{3} H_{1}} e^{-i t_{2} H_{2}} e^{-i t_{1} H_{1}}
\end{equation}

For example in the case of NMR (where orthogonal Hamiltonians are
used) we can perform universal quantum computation within three
computational steps, while the analytical solutions for the time
parameters are the trivial Euler angles. \\

\noindent \textbf{Non-Orthogonal Hamiltonians} If the Hamiltonians
are non-orthogonal i.e. $(H_1,H_2)={\rm Trace}(H_1 H_2)\ne 0$, the
number of steps--the order of generation is $n=k+2$, given by the
Lowental's criterion
\begin{equation}
\cos{(\frac{\pi}{k})}<
\frac{|(H_{1},H_{2})|}{(H_{1},H_{1})^{1/2}(H_{2},H_{2})^{1/2}}\leq\cos{(\frac{\pi}{k+1})},\quad
k\geq 2
\end{equation}
and the corresponding construction scheme is the following
\begin{equation}
  W=e^{-i t_{n} H_{1}}\ldots e^{-i
t_{3} H_{1}} e^{-i t_{2} H_{2}} e^{-i t_{1} H_{1}}
\end{equation}

For example in the case of the charge Josephson junctions where
the general Hamiltonian is $H=\frac{1}{2}
E_{c}\sigma_{z}-\frac{1}{2} E_{J}\sigma_{x}$ manipulation of the
bias energy $E_c$ which is controlled by the binary switch of gate
voltage $V_g$, provides the following non-orthogonal Hamiltonians

\begin{equation}
H_{1}= -\frac{1}{2} E_{J}\sigma_{x} \quad \mbox{and} \quad
H_{2}=\frac{1}{2} E_{c}\sigma_{z} -\frac{1}{2} E_{J}\sigma_{x}
\end{equation}

The pair $\{H_{1},H_{2}\}$ generates the $su(2)$ algebra but since
Trace$(H_{1} H_{2})\neq 0$ the whole $SU(2)$ group cannot be cover
in 3 steps.

The Lowenthal's parameter $\psi$ is

\begin{equation}
  \psi=\frac{|(H_{1},H_{2})|}{(H_{1},H_{1})^{1/2}(H_{2},H_{2})^{1/2}}
=\frac{\frac{E_{J}}{E_{c}}}{\sqrt{1+\frac{E_{J}^{2}}{E_{c}^{2}}}}=\frac{x}{\sqrt{1+x^{2}}}
\quad \mbox{where}\,\,\,x=\frac{E_{J}}{E_{c}}
\end{equation}

If $x$ is small enough, then $\psi <\cos{\frac{\pi}{3}}$ and every
element of the $SU(2)$ group
$W=w_{0}\mathbb{I}-i(w_{1}\sigma_{x}+w_{2}\sigma_{y}+w_{3}\sigma_{z})$
(one-qubit gate), can be constructed in 4 steps at most. The
corresponding analytical solutions in that case are \\

\noindent $
  t_1=-\frac{2}{E_J}\,\arctan\left({\frac{Ec(w_0w_3-w_1w_2)+\sqrt{w_2^2+w_3^2}\sqrt{E_c^2(w_0^2+w_1^2)
   -E_J^2(w_2^2+w_3^2)}}{-E_c(w_0w_2+w_1w_3)+E_J(w_2^2+w_3^2)}}\right)+\frac{4
   k_1\pi}{E_J}
$

\noindent$
  t_2=-\frac{2}{\sqrt{E_c^2+E_J^2}}\,\arctan\left({\frac{\sqrt{E_c^2+E_J^2}\sqrt{w_2^2+w_3^2}}
   {\sqrt{E_c^2(w_0^2+w_1^2)-E_J^2(w_2^2+w_3^2)}}}\right)
   +\frac{4 k_2\pi}{\sqrt{E_c^2+E_J^2}}
$

\noindent$
t_3=-\frac{2}{E_J}\,\arctan\left({\frac{Ec(w_0w_3+w_1w_2)+\sqrt{w_2^2+w_3^2}\sqrt{E_c^2(w_0^2+w_1^2)
   -E_J^2(w_2^2+w_3^2)}}{E_c(w_0w_2-w_1w_3)+E_J(w_2^2+w_3^2)}}\right)+
   \frac{4 k_3\pi}{E_J}
$

\noindent$\begin{array}{ll}
 t_4=& \frac{2}{\sqrt{E_c^2+E_J^2}}\,
     \rm{arccot}(\frac{-2(w_1w_2+w_0w_3)\sqrt{1+x^2}}{2(w_0^2+w_1^2-(w_2^2+w_3^2)\,x^2)}+
     \\ & \\
     & + \frac{\sqrt{4(w_1w_2+w_0w_3)^2(1+x^2)-4(w_0^2+w_1^2-(w_2^2+w_3^2)\,x^2)
     (2(w_0w_2-w_1w_3)\,x-(w_2^2+w_3^2)(-1+x^2))}}
     {2(w_0^2+w_1^2-(w_2^2+w_3^2)\,x^2)} )+ \\ & \\
     & + \frac{4 k_4\pi}{\sqrt{E_c^2+E_J^2}}
     \quad \mbox{where}\,\,\, k_{1}, k_{2}, k_{3}, k_{4} \in \mathbb{N} \\
\end{array}$

\subsection{Two qubit gates}

Analysis according to the "standard circuit model" requires at
least 5 Hamiltonians and 19 computational steps and and it is
performed most of the times in systems which provides
orthogonal Hamiltonians. \\

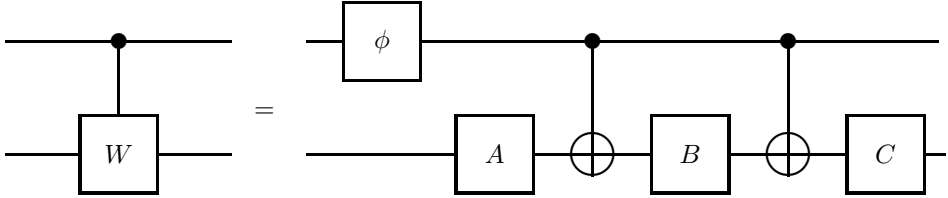
\begin{center}
\begin{picture}(13.0,2.0) \thicklines

\put(0.0,2.0){\line(1,0){3.0}} \put(0.0,0.5){\line(1,0){1.0}}
\put(1.0,0.0){\framebox(1,1){$W$}} \put(2.0,0.5){\line(1,0){1.0}}
\put(1.5,2.0){\circle*{0.2}} \put(1.5,2.0){\line(0,-1){1.0}}

\put(4.0,2.0){\line(1,0){0.5}}\put(4.5,1.5){\framebox(1,1){$\phi$}}
\put(5.5,2.0){\line(1,0){6.9}} \put(7.8,2.0){\circle*{0.2}}
\put(10.4,2.0){\circle*{0.2}}

\put(4.0,0.5){\line(1,0){2.0}}\put(6.0,0.0){\framebox(1,1){$A$}}
\put(7.0,0.5){\line(1,0){1.6}}\put(7.8,0.5){\circle{0.6}}
\put(8.6,0.0){\framebox(1,1){$B$}}\put(9.6,0.5){\line(1,0){1.6}}
\put(10.4,0.5){\circle{0.6}}\put(11.2,0.0){\framebox(1,1){$C$}}
\put(12.2,0.5){\line(1,0){0.3}}

\put(7.8,0.2){\line(0,1){1.8}}\put(10.4,0.2){\line(0,1){1.8}}

\put(3.3,1.0){$=$}
\end{picture}
\end{center}

In the case of orthogonal Hamiltonians there is the Cartan
decomposition of the $SU(2^{N})$ group \cite{Khaneja01}, directly
applied to the $SU(4)$ group, which gives analytical solutions and
was recently extended with an algorithm to realize every
$SU(2^{N})$ \cite{Earp}. The decomposition provided by
\cite{Khaneja01} requires 5 different Hamiltonians and 27
computational steps to simulate an arbitrary gate while the number
of computational steps reduces to 19 in the case of a controlled
gate. \\

Next we show the results of numerical simulations of the present
computational scheme:

\noindent \textbf{Orthogonal Hamiltonians} If the Hamiltonians are
orthogonal (e.g. Heisenberg interaction \cite{Braunstein00}, BQHD
\cite{Scarola03}, SQUIDs \cite{Mahklin00}) then with two binary
switches  providing us with 3 different Hamiltonians and within 15
computational steps we cover the $SU(4)$ group (conjecture) and
all the tested gates are successfully simulated.

For example, in a system described by a general Hamiltonian of the
form $ H=\sum_i^N \bar{B}^i(t) \hat{\sigma}^{(i)}+\sum_{i\neq
j}J_{ab}^{ij}(t) \hat{\sigma}_a^{(i)} \hat{\sigma}_b^{(j)} $
(Heisenberg interaction), only 3 Hamiltonians
\[
  H_{1}=B^{1}\, \sigma_{z}^{(1)}
\]
\[
  H_{2}=B^{2}\, \sigma_{x}^{(2)}
\]
\begin{equation}
H_3=J_{12}\,
\left(\sigma_x^{(1)}\sigma_x^{(2)}+\sigma_y^{(1)}\sigma_y^{(2)}
+\sigma_z^{(1)}\sigma_z^{(2)}\right)
\end{equation}
are sufficient for universal quantum computation in 15
computational steps
\begin{equation}\label{eq:15steps}
U=e^{H_3 t_{15}} e^{H_2 t_{14}} e^{H_1 t_{13}}
  \ldots  e^{H_3 t_{3}} e^{H_2 t_{2}}  e^{H_1 t_{1}}
\end{equation}

\noindent \textbf{Non-Orthogonal Hamiltonians} If the Hamiltonians
are non-orthogonal (charge Josephson junctions \cite{Mahklin00},
quantum dots \cite{Tanamoto00}, permanent interaction which cannot
be switched off etc) but the interaction between the qubits is
weak, then using 4 different Hamiltonians and within 15
computational steps (time parameters) a large part of the $SU(4)$
is covered and all the known important gates for quantum
computation are successfully simulated. In general, the weaker the
interaction, the larger the part of the group  covered
(more gates can be simulated). \\

\noindent a) \emph{Permanent Interaction}. If the interaction
$J_{12}$ of the previous paradigm can not be switched of then a
construction scheme with two binary switches and 3 non-orthogonal
Hamiltonians
\[
  H_{1}=B^{1}\, \sigma_{z}^{(1)}+J_{12}\,
\left(\sigma_x^{(1)}\sigma_x^{(2)}+\sigma_y^{(1)}\sigma_y^{(2)}
+\sigma_z^{(1)}\sigma_z^{(2)}\right)
\]
\[
  H_{2}=B^{2}\, \sigma_{x}^{(2)}+J_{12}\,
\left(\sigma_x^{(1)}\sigma_x^{(2)}+\sigma_y^{(1)}\sigma_y^{(2)}
+\sigma_z^{(1)}\sigma_z^{(2)}\right)
\]
\begin{equation}
H_3=J_{12}\,
\left(\sigma_x^{(1)}\sigma_x^{(2)}+\sigma_y^{(1)}\sigma_y^{(2)}
+\sigma_z^{(1)}\sigma_z^{(2)}\right)
\end{equation}
simulates all the basic gates in 15 steps
(\ref{eq:15steps}). \\

\noindent b) \emph{Charge Josephson junctions}. A system of two
identical coupled Josephson junctions is described by the
following general Hamiltonian  ${\frac{1}{2}}
 E_{c_1}\,\sigma_z^{(1)}-{\frac{1}{2}}E_{J_1}\,\sigma_x^{(1)}+
 {\frac{1}{2}}E_{c_2}\,\sigma_z^{(2)}-{\frac{1}{2}}E_{J_2}\,\sigma_x^{(2)}-
 {\frac{1}{2}}E_L\,\sigma_y^{(1)}\,\sigma_y^{(2)}$. Manipulation
 of 3 binary switches of the system provides  the following
 4 non-orthogonal Hamiltonians
\[
 H_{1} = {\frac{1}{2}} E_{c}\,(
 \sigma_z^{(1)}+\sigma_z^{(2)})
 -{\frac{1}{2}} E_J\,( \sigma_x^{(1)}+\sigma_x^{(2)} )
\]
\[
 H_{2} = -{\frac{1}{2}} E_J\,(
 \sigma_x^{(1)}+\sigma_x^{(2)} )-{\frac{1}{2}}
 E_L\,\sigma_y^{(1)}\sigma_y^{(2)}
\]
\[
 H_{3} = {\frac{1}{2}}
 E_{c}\,\sigma_z^{(2)}-{\frac{1}{2}} E_J\,(
 \sigma_x^{(1)}+\sigma_x^{(2)})
\]
\begin{equation}
 H_{4} = {\frac{1}{2}}
 E_{c}\,\sigma_z^{(1)}-{\frac{1}{2}} E_J\,(
 \sigma_x^{(1)}+\sigma_x^{(2)})
\end{equation}
and all basic gates are simulated in 15 steps (\ref{eq:15steps}).
\\

The efficiency of our simulation is defined by a test function,
$f_{test}$. It is a function of 15 time variables
\begin{equation}\label{eq:f_test}
 f_{test}(t_{1},\ldots,t_{15})=\sum_{i,j=1}^4
 |(U_{\rm gate})_{ij}-(U(t_{1},\ldots,t_{15}))_{ij}|^2=
 || U_{\rm gate} - U ||^2
\end{equation}
In our numerical simulations $f_{test}$ usually attains values of
$10^{-8}$ or less. Taking into account more decimal digits in the
approximation of the time parameters results to a further decrease
of its value. Gates that have been tested numerically are all the
important two-qubit gates for quantum computation such as the CNOT
gate, the SWAP gate, the Quantum Fourier Transform gate for two
qubits, several controlled gates etc. The ratio of the values of
the external switches tuning amplitudes over the magnitude of the
interaction is not in the area of hard pulses. Numerical results
are available upon request from the authors.

\section*{Acknowledgements}
The work of K.Ch. Ch. was supported by Herakleitos Research
Scholarships (21866). The authors would like to thank I. Jex and
G. Chadjitaskos of the Czech Technical University in Prague, D.
Angelakis and S. G. Schirmer  of the Centre for Quantum
Computation (DAMTP-University of Cambridge) for fruitful
discussions.

\end{document}